\documentclass[10pt,journal,onecolumn]{IEEEtran}
\usepackage{amsfonts}
\usepackage{amsmath}
\usepackage{amssymb}
\usepackage{bm}
\usepackage{xcolor,graphicx,float}
\usepackage{color, soul}
\usepackage{stfloats}
\usepackage[numbers,sort&compress]{natbib}
\usepackage[amsmath,thmmarks]{ntheorem}
\usepackage{theorem}
\usepackage{algorithm}
\usepackage{algorithmic}
\usepackage{graphicx}
\usepackage{subfigure}
\usepackage{pict2e}
\usepackage{tikz}
\usetikzlibrary{fit,positioning}

\theoremheaderfont{\sc}\theorembodyfont{\upshape}
\theoremstyle{nonumberplain}
\theoremseparator{}
\theoremsymbol{\rule{1ex}{1ex}}

\begin{document}
\title{Vector Approximate Message Passing Algorithm for Structured Perturbed Sensing Matrix}

\author{Jiang~Zhu, Qi Zhang, Xiangming Meng and Zhiwei Xu}
\maketitle
\begin{abstract}
In this paper, we consider a general form of noisy compressive sensing (CS) where the sensing matrix is not precisely known. Such cases exist when there are imperfections or unknown calibration parameters during the measurement process. Particularly, the sensing matrix may have some structure, which makes the perturbation follow a fixed pattern. While previous work has focused on extending the approximate message passing (AMP) and LASSO algorithm to deal with the independent and identically distributed (i.i.d.) perturbation, we propose the robust variant vector approximate message passing (VAMP) algorithm with the perturbation being structured, based on the recent VAMP algorithm. The performance of the robust version of VAMP is demonstrated numerically.
\end{abstract}
%
\section{Introduction}\label{sec:intro}
Compressed Sensing (CS) aims to reconstruct an $N$-dimensional sparse signal from $M$ underdetermined linear measurements ${\mathbf y}={\mathbf A}{\mathbf x}+{\mathbf w}$, where $M<N$ and $\mathbf w$ is additive noise. It has been shown that in the absence of noise, perfect reconstruction is possible given that the signal is exactly $K$ sparse and the measurement matrix satisfies certain properties (e.g., restricted isometry, spark, null space). In practical applications, the measurement matrix $\mathbf A$ may not be known exactly due to, e.g., model mismatch, imperfect calibration and imperfections in the signal acquisition hardware. Consequently, several works have studied the recovery algorithm and performance bounds for the general signal with independent and identically distributed (i.i.d.) perturbation \cite{Compare_model_unc}. In addition, the measurement matrix uncertainty in quantized settings has also been studied \cite{zhu1}.

For the sparse signal recovery under i.i.d. perturbation, the recovery performance of algorithms such as basis pursuit (BP) and orthogonal matching pursuit (OMP) algorithm are analyzed \cite{Herman1, Gu1}. While the above works study the effect of perturbation on established algorithms, there also exist some algorithms which take the measurement matrix uncertainty into account. In \cite{Zhu1}, the Sparsity-cognizant Total Least Squares (S-TLS) approach is developed. A modified version of the Dantzig selector dealing with the matrix uncertainty is proposed in \cite{Rosen}. To taking the structure of perturbation into account, a weighted S-TLS (WS-TLS) is proposed, and numerical results demonstrate that WS-TLS performs significantly better than S-TLS \cite{Zhu1}.

Approximate message passing (AMP) algorithm is a popular method for performing high dimensional inference, due to its low computational complexity and good performance \cite{Donoho1}. In \cite{Rangan}, a generalized AMP (GAMP) algorithm is proposed to cope with the generalized linear model \cite{Rangan}. Since then, AMP and GAMP algorithm has been applied in various signal processing applications, such as data detection and channel estimation \cite{Schniter1}. Recently, orthogonal AMP \cite{OAMP} and vector AMP (VAMP) algorithms \cite{VAMP} are proposed, which can deal with a larger ensemble of measurement matrix set, compared to the AMP algorithm. Given that some statistical parameters are unknown, expectation maximization approximate message passing (EM-AMP) and expectation maximization vector approximate message passing (EM-VAMP) are proposed to jointly recover the unknown signal and learn the statistical parameters \cite{Jeremy, Fletcher1}.

In \cite{Parker1}, an AMP algorithm is extended to deal with the sparse signal recovery problem under matrix uncertainty. The perturbation is treated as an additive white Gaussian noise, and the matrix uncertainty GAMP (MU-GAMP) is proposed. Provided that the perturbation has some additional structure, an alternating MU-GAMP is proposed to jointly estimate the measurement matrix and signal, in contrast with this paper where the structured perturbation is also treated as the random variables. In \cite{Krzakala1}, the robust approximate message passing algorithm is proposed, and the mean square error of the Bayes-optimal reconstruction of sparse signals under matrix uncertainty is calculated via replica method.
%
%

In this paper, we consider a kind of general structured perturbation. This structure arises because the sensing matrix has known structure such that its elements can not be chosen arbitrarily. For example, in signal and communication problems, the convolving operation between channel and data can be reformulated as a linear regression problem. For the zero boundary conditions, the sensing matrix has a Toeplitz structure, while a circulant structure appears for periodic boundary conditions \cite{DIM}. As a result, the structure of model uncertainty has to be taken into account to improve the reconstruction performance. Since the equivalent noise (perturbation plus additive noise) is coloured and related to the unknown signal, in contrast with the white Gaussian noise in \cite{Parker1}, conventional AMP and VAMP algorithm can not be applied in this scenario. Here we propose to approximate the likelihood function in each iteration, and numerical results demonstrate the effectiveness of the proposed method.

\section{Algorithm}
\label{sec:format}

The mathematical model we consider in this paper is \cite{Beck1}
\begin{align}\label{org1}
{\mathbf y}=\left({\mathbf A}+{\sum_{i=1}^qe_i{\mathbf E}_i}\right){\mathbf x}+{\mathbf w}.
\end{align}
where we assume that ${\mathbf y}\in {\mathbb R}^M$, ${\mathbf A}\in {\mathbb R}^{M\times N}$ denotes the random known sensing matrix and $\|{\mathbf A}\|_{\rm F}^2=N$, where $\|\cdot\|_{\rm F}$ denotes the Frobenius norm, ${\mathbf E}_i\in {\mathbb R}^{M\times N}$ denotes the known structure of the perturbation, $e_i,~i=1,\cdots,q$ are i.i.d. random variables and satisfying $e_i\sim {\mathcal N}(e_i;0,\gamma_e^{-1})$ \footnote{Here ${\mathcal N}(e_i;0,\gamma_e^{-1})$ means that $e_i$ follows Gaussian distribution with mean zero and variance $\gamma_e^{-1}$. Sometimes we use ${\mathcal N}(0,\gamma_e^{-1})$ instead when the random variable is clear.}, ${\mathbf x}\in {\mathbb R}^N$. The prior distribution of signal $\mathbf x$ follows ${\mathbf x}\sim \prod\limits_{i=1}^{N}p(x_i)$, where $p(x_i)$ is a sparsity-inducing prior, ${\mathbf w}\sim {\mathcal N}({\mathbf 0},\gamma_w^{-1}{\mathbf I}_M)$. Note that \cite{Parker1} and \cite{Beck1} study model (\ref{org1}). However, \cite{Parker1} treats $\{a_i\}_{i=1}^{q}$ as unknown deterministic parameters in contrast to \cite{Beck1} as random parameters, which correspond to two classical ways to model measurement uncertainty. As shown in \cite{Beck1}, the strategy of modeling measurement uncertainty as random parameters yields accurate results. The drawback is that one needs to estimate the statistics of the random parameters. Compared to \cite{Beck1} which assumes an unknown deterministic vector $\mathbf x$, this paper enforces prior distribution of $\mathbf x$. The perturbation model in (\ref{org1}) is very general and we list some specific structure of perturbation as follows:
\begin{itemize}
  \item i.i.d perturbation, where the perturbation takes the form $\sum_{i=1}^M{\sum_{j=1}^Ne_{ij}{\mathbf E}_{ij}}$, $e_{ij}\sim {\mathcal N}(0,\gamma_e^{-1})$ and ${\mathbf E}_{ij}$ is a all zero matrix except that the $(i,j)$-th element is one.
  \item Matrix-restricted structured perturbation  where the perturbation takes the form ${\mathbf D}{\mathbf E}{\mathbf C}$ with $\mathbf D$ and $\mathbf C$ being known matrices. This structure can model the scenario in which the coefficients of the sensing matrix have unequal uncertainties, as shown in \cite{Beck1}.
  \item Circulant structure perturbation. Here the $N\times N$ circulant matrix ${\mathbf A}$ is of the form
  \begin{align}\label{circ}
  {\mathbf A}=\left[
                \begin{array}{cccc}
                  a_1 & a_2 & \cdots & a_N \\
                  a_N & a_1 & \cdots & a_{N-1} \\
                  \vdots & \vdots & \vdots & \vdots \\
                  a_2 & a_3 & \cdots & a_1 \\
                \end{array}
              \right].
  \end{align}
As a result, the perturbation also takes this form \cite{Beck1}.
\end{itemize}


By defining ${\mathbf z}={\sum_{i=1}^qe_i{\mathbf E}_i}{\mathbf x}+{\mathbf w}$, model (\ref{org1}) is equivalent to
\begin{align}\label{eq_model}
{\mathbf y}={\mathbf A}{\mathbf x}+{\mathbf z},
\end{align}
where ${\mathbf z}\sim {\mathcal N}({\mathbf 0},{\sum_{i=1}^q}\gamma_e^{-1}{\mathbf E}_i{\mathbf x}{\mathbf x}^{\rm T}{\mathbf E}_i^{\rm T}+\gamma_w^{-1}{\mathbf I}_M)\triangleq {\mathcal N}({\mathbf 0},{\boldsymbol \Gamma}({\mathbf x}))$.
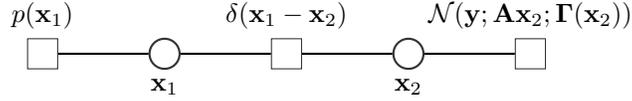
\begin{figure}
\centering
\begin{tikzpicture}
\tikzstyle{main}=[circle, minimum size = 4mm, thick, draw =black!80, node distance = 12mm]
\tikzstyle{connect}=[-latex, thick]
\tikzstyle{box}=[rectangle, minimum size = 4mm, draw=black!80, node distance = 12mm]
  \node[box, fill = white!100] (px1) [label=above:$p({\mathbf x}_1)$] { };
  \node[main] (x1) [right=of px1,label=below:${\mathbf x}_1$] { };
  \node[box] (delta1) [right=of x1,label=above:$\delta({\mathbf x}_1-{\mathbf x}_2)$] {};
  \node[main] (x2) [right=of delta1,label=below:${\mathbf x}_2$] { };
  \node[box, fill = white!100] (y) [right=of x2,label=above:${\mathcal N}({\mathbf y};{\mathbf A}{\mathbf x}_2;{\boldsymbol \Gamma}({\mathbf x}_2))$] { };
\draw[thick] (px1.east) -- (x1.west);
\draw[thick] (x1.east) -- (delta1.west);
\draw[thick] (delta1.east) -- (x2.west);
\draw[thick]  (x2.east) -- (y.west);
\end{tikzpicture}
\caption{The factor graph used for the derivation of the robust VAMP algorithm. The circles represent variable nodes and the squares represent factor nodes from (\ref{split-fac}).} \label{fig:bvamp1}
\end{figure}
In the following text, we introduce the VAMP briefly \footnote{For the detailed derivation of VAMP utilizing expectation propagation, please refer to \cite{VAMP}.}. We start with the joint probability density function of $\mathbf x$ and $\mathbf y$ as
\begin{align}
p({\mathbf y},{\mathbf x})=p({\mathbf x})p({\mathbf y}|{\mathbf x})=p({\mathbf x}){\mathcal N}({\mathbf y};{\mathbf A}{\mathbf x},{\boldsymbol \Gamma}({\mathbf x})).
\end{align}
By splitting x into two identical variables ${\mathbf x}_1$ and ${\mathbf x}_2$, we obtain an equivalent factorization
\begin{align}\label{split-fac}
p({\mathbf y},{\mathbf x}_1,{\mathbf x}_2)=p({\mathbf x}_1)\delta({\mathbf x}_1-{\mathbf x}_2){\mathcal N}({\mathbf y};{\mathbf A}{\mathbf x}_2,{\boldsymbol \Gamma}({\mathbf x}_2)).
\end{align}
The factor graph corresponding to the above factorization (\ref{split-fac}) is presented in Fig. \ref{fig:bvamp1}. We then pass messages on this factor graph. We initialize the message of the factor node $\delta({\mathbf x}_1-{\mathbf x}_2)$ to the variable node ${\mathbf x}_1$ with $\mu_{\delta\rightarrow {\mathbf x}_1}({\mathbf x}_1)={\mathcal N}({\mathbf x}_1;{\mathbf r}_{1k},\gamma_{1k}^{-1}{\mathbf I}_N)$ where $k=0$. Combing the factor node $p({\mathbf x}_1)$, the sum product (SP) belief on variable node ${\mathbf x}_1$ is
\begin{align}
b_{\rm sp}({\mathbf x}_1)\propto p({\mathbf x}_1){\mathcal N}({\mathbf x}_1;{\mathbf r}_{1k},\gamma_{1k}^{-1}{\mathbf I}_N).
\end{align}
where $\propto$ means proportional to. We calculate the posterior means and variances as
\begin{subequations}\label{xhat1k}
\begin{align}
&\hat{\mathbf x}_{1k}={\rm E}[{\mathbf x}_1|b_{\rm sp}({\mathbf x}_1)],\\
&\eta_{1k}^{-1}=<{\rm diag}({\rm Cov}[{\mathbf x}_1|b_{\rm sp}({\mathbf x}_1)])>,
\end{align}
\end{subequations}
where $<{\mathbf x}>=(\sum\limits_{i=1}^Nx_i)/N$, ${\rm Cov}[\cdot|b_{\rm sp}({\mathbf x}_1)]$ is the covariance matrix with respect to the belief estimate $b_{\rm sp}({\mathbf x}_1)$ and ${\rm diag}({\mathbf A})$ returns a column vector whose elements are the main diagonal of ${\mathbf A}$. Exploiting the expectation propagation, the above belief $b_{\rm sp}({\mathbf x}_1)$ is approximated as a Gaussian distribution $b_{\rm app}({\mathbf x}_1)$ given by
\begin{align}
b_{\rm app}({\mathbf x}_1)={\mathcal N}({\mathbf x}_1;\hat{\mathbf x}_{1k},\eta_{1k}^{-1}{\mathbf I}_N).
\end{align}
Then we calculate the message from the variable node ${\mathbf x}_1$ to the factor node $\delta({\mathbf x}_1-{\mathbf x}_2)$, which is the ratio of the most recent approximate belief $b_{\rm app}({\mathbf x}_1)$ to the most recent message from $\delta({\mathbf x}_1-{\mathbf x}_2)$ to ${\mathbf x}_1$, i.e.,
\begin{align}
&\mu_{{\mathbf x}_1\rightarrow\delta}={\mathcal N}({\mathbf x}_1;{\mathbf r}_{2k},\gamma_{2k}{\mathbf I}_N)\notag\\
&\propto {\mathcal N}({\mathbf x}_1;\hat{\mathbf x}_{1k},\eta_{1k}^{-1}{\mathbf I}_N)/{\mathcal N}({\mathbf x}_1;{\mathbf r}_{1k},\gamma_{1k}^{-1}{\mathbf I}_N),
\end{align}
where ${\mathbf r}_{2k}$ and $\gamma_{2k}$ are calculated according to line 5 in Algorithm 1. For the factor node $\delta({\mathbf x}_1-{\mathbf x}_2)$, the message from the factor node $\delta({\mathbf x}_1-{\mathbf x}_2)$ to the variable node ${\mathbf x}_2$ can be calculated directly as $\mu_{\delta\rightarrow{\mathbf x}_2}({\mathbf x}_2)={\mathcal N}({\mathbf x}_2;{\mathbf r}_{2k},\gamma_{2k}{\mathbf I}_N)$ which can be viewed as the prior of the variable node ${\mathbf x}_2$. For the rightmost factor node ${\mathcal N}({\mathbf y};{\mathbf A}{\mathbf x}_2;{\boldsymbol \Gamma}({\mathbf x}_2))$, its covariance matrix depends on the unknown $\mathbf x$. As a result, we approximate ${\boldsymbol \Gamma}({\mathbf x}_2)$ as
\begin{align}
&{\boldsymbol \Gamma}({\mathbf x}_2)\approx {\rm E}_{{\mathbf x}_2\sim {\mathcal N}({\mathbf r}_{2k},\gamma_{2k}^{-1}{\mathbf I})}\left[{\boldsymbol \Gamma}({\mathbf x}_{2k})\right]\notag\\
&={\sum_{i=1}^q}\gamma_a^{-1}{\mathbf A}_i({\mathbf r}_{2k}{\mathbf r}_{2k}^{\rm T}+\gamma_{2k}^{-1}{\mathbf I}){\mathbf A}_i^{\rm T}+\gamma_w^{-1}{\mathbf I}_m\triangleq {\boldsymbol \Gamma}_{2k}.
\end{align}
As a consequence, we obtain an approximate model with the likelihood ${\mathcal N}({\mathbf y}_{2k};{\mathbf A}_{2k}{\mathbf x}_2; \gamma_{w,2k}^{-1}{\mathbf I}_M)
$, where
\begin{align}
{\mathbf y}_{2k} = \gamma_{w,2k}^{-\frac{1}{2}}{\boldsymbol \Gamma}_{2k}^{-\frac{1}{2}}{\mathbf y},\label{y2k}\\
{\mathbf A}_{2k} =  \gamma_{w,2k}^{-\frac{1}{2}}{\boldsymbol \Gamma}_{2k}^{-\frac{1}{2}}{\mathbf A}\label{A2k},
\end{align}
where $\gamma_{w,2k}^{-\frac{1}{2}}$ is to ensure $\|{\mathbf A}_{2k}\|_{\rm F}^2=N$ and $\gamma_{w,2k}=\|{\boldsymbol \Gamma}_{2k}^{-\frac{1}{2}}{\mathbf A}\|_{\rm F}^2/N$. With such an approximation, the SP belief on variable ${\mathbf x}_2$ is
\begin{align}
b_{\rm sp}({\mathbf x}_2)\propto {\mathcal N}({\mathbf y}_{2k};{\mathbf A}_{2k}{\mathbf x}_2; \gamma_{w,2k}^{-1}{\mathbf I}_M){\mathcal N}({\mathbf x}_2,{\mathbf r}_{2k},\gamma_{2k}{\mathbf I}_N).
\end{align}
Utilizing the expectation propagation, the SP belief $b_{\rm sp}({\mathbf x}_2)$ on variable ${\mathbf x}_2$ can be further approximated as
\begin{align}
b_{\rm app}({\mathbf x}_2)={\mathcal N}({\mathbf x}_2;\hat{\mathbf x}_{2k},\eta_{2k}^{-1}{\mathbf I}_N),
\end{align}
where
\begin{subequations}
\begin{align}\label{LMMSE}
&\hat{\mathbf x}_{2k}=(\gamma_{w,2k}{\mathbf A}_{2k}^{\rm T}{\mathbf A}_{2k}+\gamma_{2k}{\mathbf I})^{-1}
(\gamma_{w,2k}{\mathbf A}_{2k}^{\rm T}{\mathbf y}_{2k}+\gamma_{2k}{\mathbf r}_{2k}),\notag\\
&\eta_{2k}^{-1}=\frac{1}{N}{\rm Tr}\left[(\gamma_{w,2k}{\mathbf A}_{2k}^{\rm T}{\mathbf A}_{2k}+\gamma_{2k}{\mathbf I})^{-1}\right].
\end{align}
\end{subequations}
We then obtain the message from the variable node ${\mathbf x}_2$ to the factor node $\delta({\mathbf x}_1-{\mathbf x}_2)$ with
\begin{align}
&\mu_{{\mathbf x}_2\rightarrow \delta}({\mathbf x}_2)\propto b_{\rm app}({\mathbf x}_2)/{\mathcal N}({\mathbf x}_2,{\mathbf r}_{2k},\gamma_{2k}{\mathbf I}_N)\notag\\
=&{\mathcal N}({\mathbf x}_2;{\mathbf r}_{1,k+1},\gamma_{1,k+1}^{-1}{\mathbf I}_n),
\end{align}
where ${\mathbf r}_{1,k+1}$ and $\gamma_{1,k+1}^{-1}$ are given in line 10 in Algorithm 1. Similarly, we calculate the message from the variable node ${\mathbf x}_2$ to the factor node $\delta({\mathbf x}_1-{\mathbf x}_2)$ as
\begin{align}
&\mu_{\delta\rightarrow{\mathbf x}_1}({\mathbf x}_1)=\mu_{{\mathbf x}_2\rightarrow\delta}({\mathbf x}_2),
\end{align}
which closes the loop of the proposed VAMP algorithm and is shown in Algorithm 1.
\begin{algorithm}[h]
\caption{Vector AMP in perturbed setting}
\begin{algorithmic}[1]
\STATE Initialize ${\mathbf r}_{10}$ and ${\gamma}_{10}\geq 0$ and set the maximum number of iterations $K_{\rm it}$;\
\FOR {$k=0,1,\cdots,K_{\rm it}$ }
\STATE //~Denoising
\STATE Calculate $\hat{\mathbf x}_{1k}$ and ${\eta}_{1k}$ according to (\ref{xhat1k}).
\STATE ${\gamma}_{2k}={\eta}_{1k}-\gamma_{1k}$,~${\mathbf r}_{2k}=({\eta}_{1k}\hat{\mathbf x}_{1k}-{\gamma}_{1k}{\mathbf r}_{1k})/{\gamma}_{2k}$\
\STATE //~Whitening
\STATE Approximate ${\boldsymbol \Gamma}({\mathbf x}_2)$ with ${\boldsymbol \Gamma}_{2k}$ and obtain the equivalent ${\mathbf y}_{2k}$ (\ref{y2k}), ${\mathbf A}_{2k}$ (\ref{A2k}) and $\gamma_{w,2k}$.
\STATE //~LMMSE estimation
\STATE Calculate $\hat{\mathbf x}_{2k}$ and ${\eta}_{2k}$ according to (\ref{LMMSE}).\
\STATE ${\gamma}_{1,k+1}={\eta}_{2k}-\gamma_{2k}$,\\${\mathbf r}_{1,k+1}=({\eta}_{2k}\hat{\mathbf x}_{2k}-{\gamma}_{2k}{\mathbf r}_{2k})/{\gamma}_{1,k+1}$\
\ENDFOR
\STATE Return $\hat{\mathbf x}_{1K_{\rm it}}$.
\end{algorithmic}
\end{algorithm}

Now we discuss the computation complexity of Algorithm 1. For the VAMP presented in \cite{VAMP}, the main computational burden lies in the SVD of the sensing matrix, which performs only once. For Algorithm 1, the main additional computational burden lies in line 7, which involves in calculating the eigenvalue decomposition of ${\boldsymbol \Gamma}_{2k}$ and the singular value decomposition of ${\mathbf A}_{2k}$ for each iteration. For some cases, the computation complexity of Algorithm 1 is comparable to that of VAMP. Given that the model is ${\mathbf y}=({\mathbf A}+{\mathbf E}){\mathbf x}+{\mathbf n}$ and the elements of perturbation ${\mathbf E}$ are i.i.d., where $E_{ij}\sim {\mathcal N}(0,\gamma_e^{-1})$, one can see that ${\boldsymbol \Gamma}({\mathbf x})=(\gamma_w^{-1}+\gamma_e^{-1}\|{\mathbf x}\|_2^2){\mathbf I}_M$ and the whitening operation in line 7 is unnecessary.

\section{Numerical Results}
\label{sec:pagestyle}

In this section, numerical results are performed to verify the effectiveness of the proposed robust VAMP. The performance of the following algorithms are evaluated:
\begin{itemize}
  \item The AMP-oracle algorithm with precisely known sensing matrix
  \item The PI-AMP algorithm which does not take the perturbation into account.
  \item The MU-GAMP algorithm presented in \cite{Parker1}.
  \item The VAMP-oracle algorithm with precisely known sensing matrix
  \item The VAMP-PI algorithm which ignores perturbation
  \item The VAMP-PC algorithm shown in Algorithm 1 which considers model perturbation.
\end{itemize}
In the numerical simulation, we assume a Bernoulli Gaussian prior, i.e., $p(x_i)=(1-\rho)\delta(x_i)+\rho{\mathcal N}(x_i,\mu_x,\sigma_x^2)$, where $\rho=0.2$, $\mu_x=0$ and $\sigma_x^2=1$. For the first two numerical experiments, the elements of matrix $\mathbf A$ are i.i.d. drawn from Gaussian distribution. We assume that each deterministic ${\mathbf E}_i$ is also drawn from Gaussian distribution, and we set $M=0.5N$ and $q=N$. The normalized mean square error (NMSE) is defined as ${\rm NMSE}(\hat{\mathbf x})=10\log \frac{\|{\mathbf x}-\hat{\mathbf x}\|_2^2}{\|{\mathbf x}\|_2^2}$, where ${\mathbf x}$ denotes the true value. We also define ${\rm SNR}_{\rm w}\triangleq 10\log \frac{\|\mathbf {Ax}\|^2}{\|{\mathbf w}\|^2}$ and ${\rm SNR}_{\rm e}\triangleq 10\log \frac{\|{\sum_{i=1}^qe_i{\mathbf E}_i{\mathbf x}}\|^2}{\|{\mathbf w}\|^2}$. The maximum number of iterations is $K_{\rm it}=60$.

\begin{figure}[htb]
  \centering
  \centerline{\includegraphics[width=8cm]{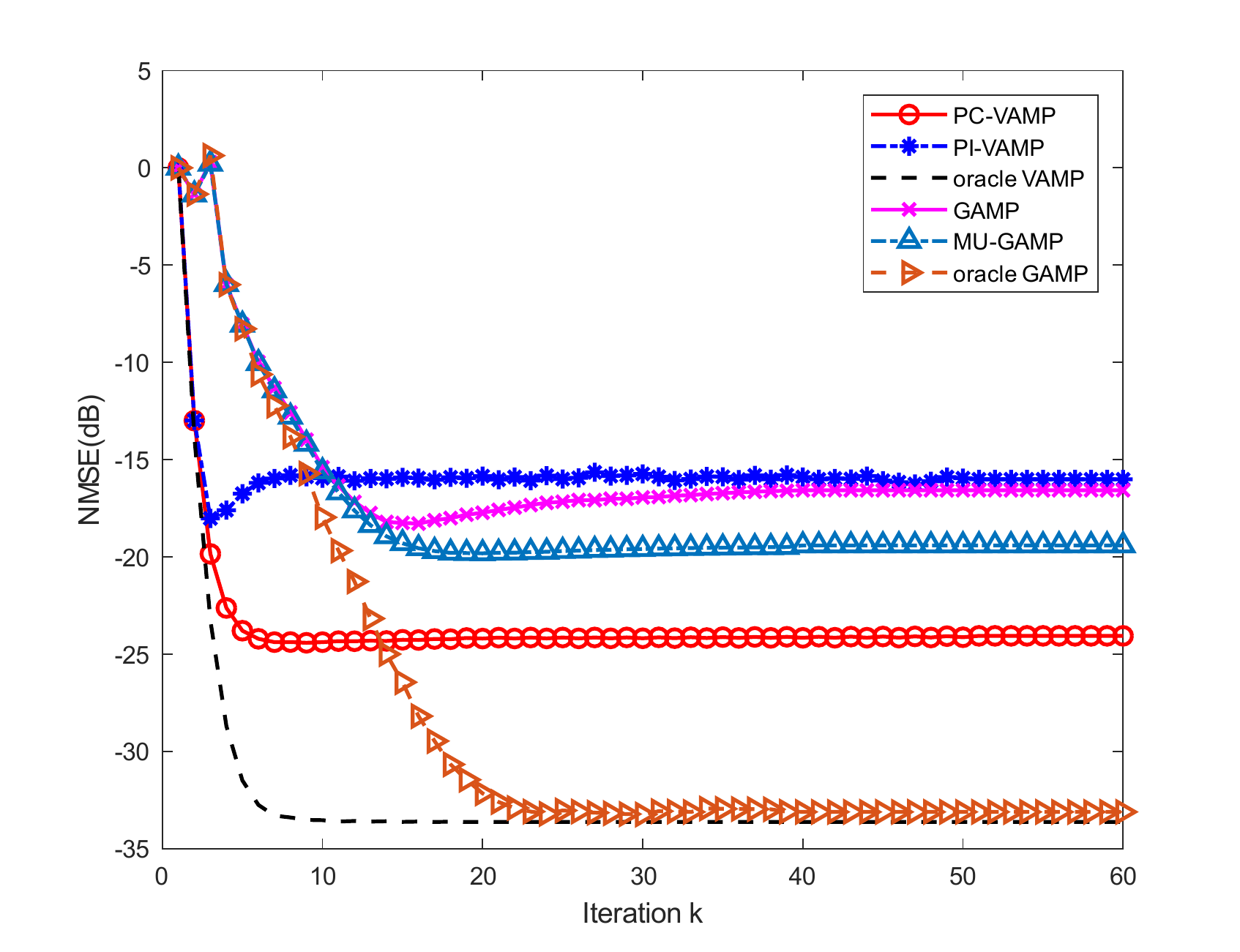}}
\caption{NMSE versus algorithm iteration in a single realization}\label{Fig1}
\end{figure}

In the first numerical simulation, the NMSE versus iteration is presented. We set $\rm{SNR_w} = 30dB$ and $\rm{SNR_e} = 20dB$. From Fig. \ref{Fig1}, one can see that the oracle VAMP and GAMP algorithm achieves the lowest NMSE, and the oracle VAMP achieves the fastest speed of convergence. For unknown structured perturbation, PC-VAMP works better than MU-GAMP.

\begin{figure}
  \centering
  \includegraphics[width=8cm]{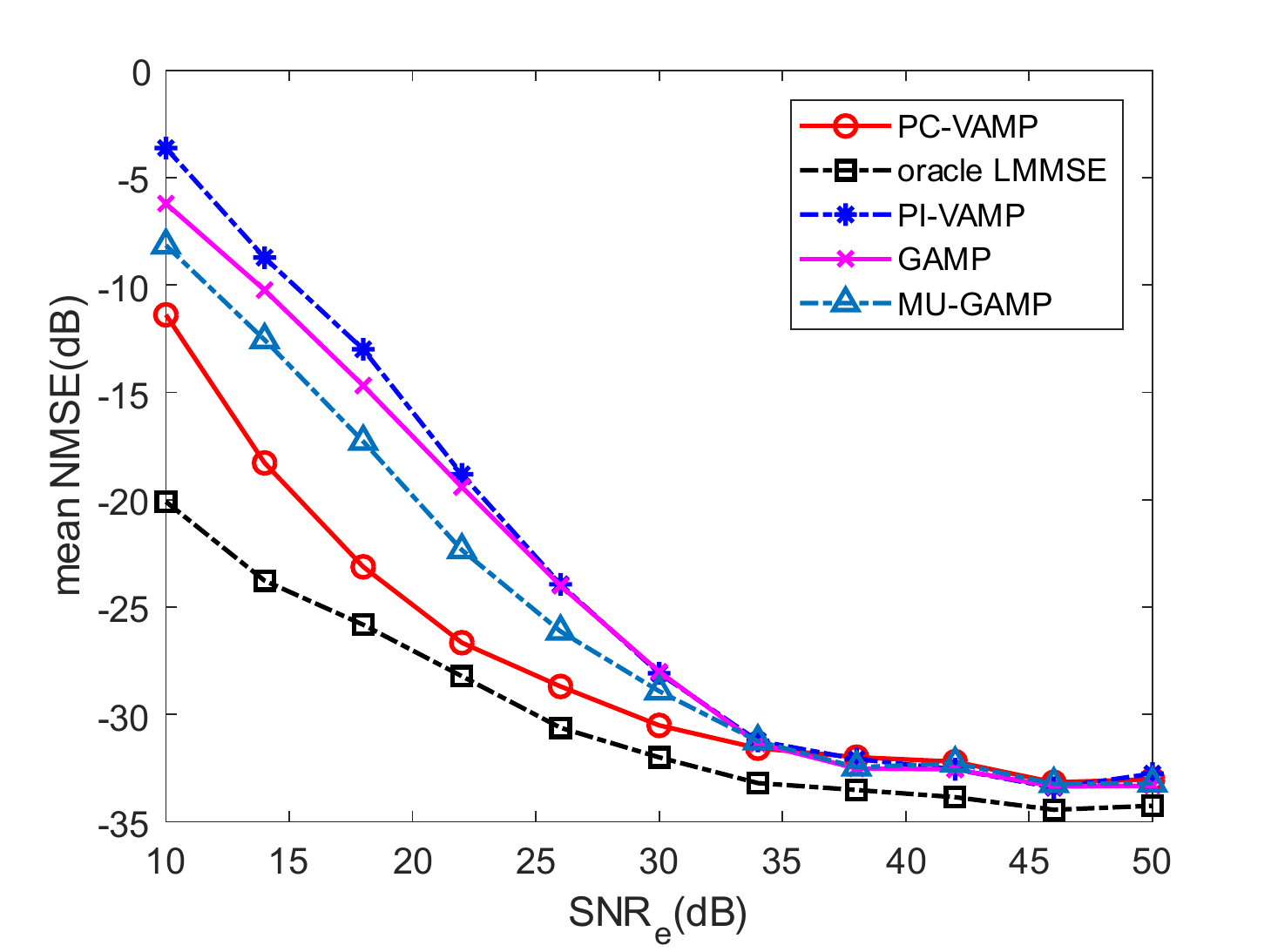}\\
  \caption{Mean NMSE versus $\rm{SNR_e}$. The reported NMSE is averaged over $50$ realizations.}\label{NMSEvgammaa}
\end{figure}

In the second simulation, all the parameters are the same as that in the first simulation and $\rm{SNR_w} = 30dB$. In Fig. \ref{NMSEvgammaa}, we see that there exists obvious performance gap between PI-VAMP algorithm and MU-GAMP given ${\rm SNR}_e\leq 30{\rm dB}$. Compared to the MU-GAMP algorithm, PC-VAMP algorithm works better. When the perturbation is small such that ${\rm SNR}_e\geq 35{\rm dB}$, the performances of all the AMP and VAMP algorithms are similar.

\begin{figure}p
  \centering
  \includegraphics[width=8cm]{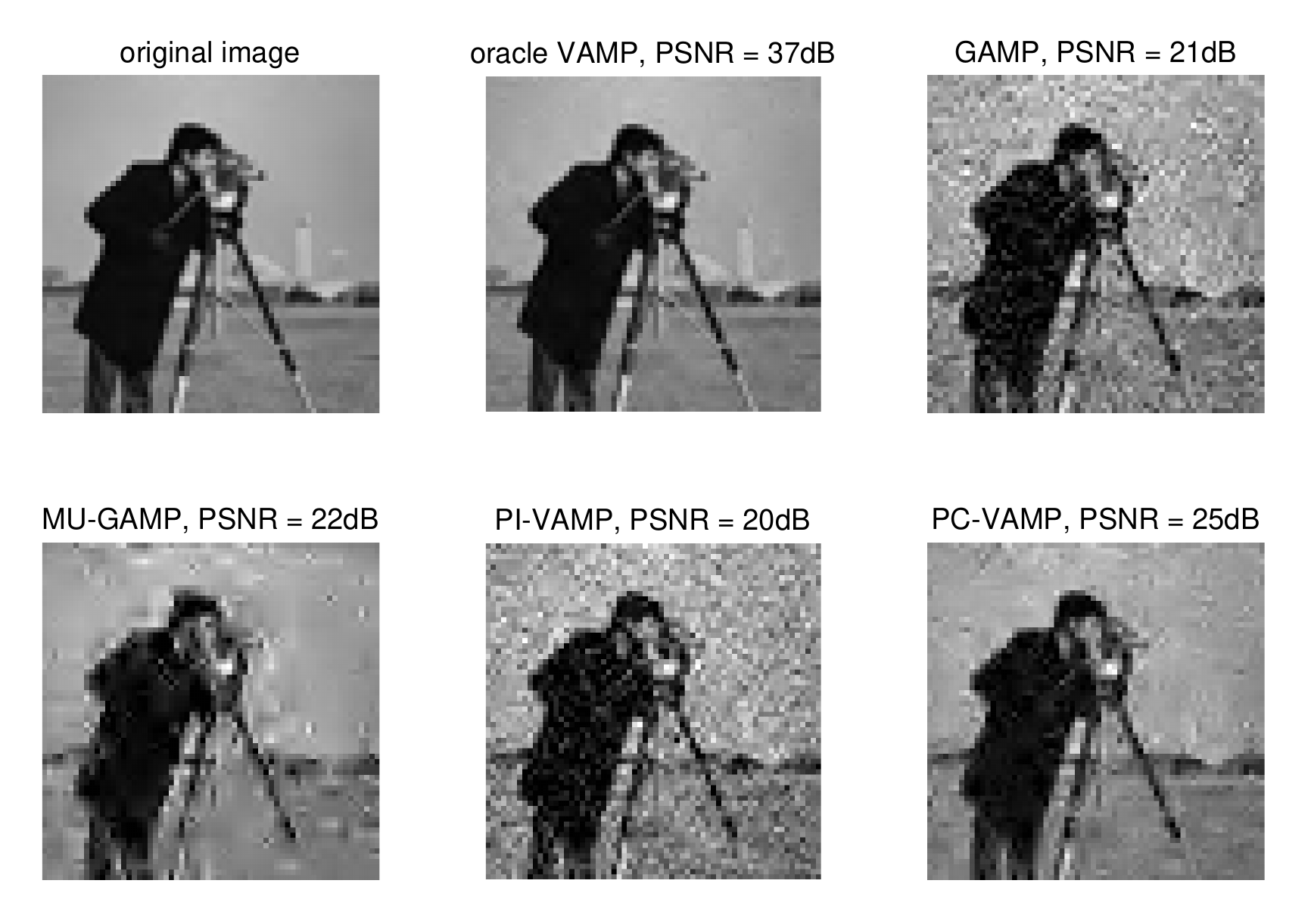}\\
  \caption{$64\times 64$ image recovery results.}\label{came}
\end{figure}

The last experiment investigates the performance of PC-VAMP algorithm for real image recovery. We threshold the wavelet coefficients such that the sparsity is $\rho=0.4133$. We set $\mu_x=3.0\times 10^{-4}$, $\sigma_x^2=1.7\times10^{-2}$. We use a $N\times N$ circulant matrix (\ref{circ}), set $a_i=0.3^i,~i=0,\cdots,N-1$ and $N=64^2$. The perturbation also has the circulant structure. We use a random matrix to compress the observations such that the measurement ratio is $0.9$. For this compressed observation model, we set $\rm{SNR_w}=40dB$ and $\rm{SNR_e}=20dB$.  From Fig. \ref{came}, it can be seen that PC-VAMP yields the best recovery results with the perturbation being unknown, and the PSNR is $25{\rm dB}$.
\section{Conclusion}
\label{sec:typestyle}
In this paper, we propose a matrix-uncertainty extension of the VAMP algorithm, when some structured perturbation is added on the sensing matrix. By iteratively approximating the original likelihood function with constant covariance matrix, we obtain a modified VAMP algorithm. Numerical results demonstrate the effectiveness of the proposed algorithm.
\bibliographystyle{IEEEbib}
\bibliography{strings,refs}

\end{document}